\documentstyle[preprint,prd,aps]{revtex}

\advance\hoffset by -3mm  
\advance\voffset by  8mm  

\begin{document}

\tolerance=10000
\bibliographystyle{prsty}

\draft

\author{C.~Q.~Geng$^{a}$, I.~J.~Hsu$^{b}$ and Y.~C.~Lin$^{b}$}

\address{
$^a$Department of Physics, National Tsing Hua University \\
Hsinchu, Taiwan 30043, Republic of China \\
and \\
$^b$Department of Physics, National Central University\\
Chung-Li, Taiwan 32054, Republic of China}

\title{
Study of Long Distance Contributions to $K\rightarrow n\pi\nu\bar{\nu}$}
\maketitle

\begin{abstract}
We calculate long distance contributions to
$K\rightarrow\pi\nu\bar{\nu}\,,\ \pi\pi\nu\bar{\nu}$, and
$\pi\pi\pi\nu\bar{\nu}$  modes within the framework of chiral
perturbation theory.
We find that these contributions to decay rates of
$K\rightarrow \pi\nu\bar{\nu}$ and $K\rightarrow \pi\pi\nu\bar{\nu}$ 
in the chiral logarithmic approximation
are at least seven orders of magnitude suppressed relative to
those from the short distance parts.
The long distance effects in this class of decays are therefore negligible.

\end{abstract}


\section{Introduction}
The $K$ decays receive contributions
from both long and short
distance effect. The long distance one is the sum of all sorts of
nonperturbative effects such as hadron formation, symmetry breaking, {\it etc}.
These effects realize themselves as the coefficients in the chiral
Lagrangian and can only be determined by experimental fits. The short
distance contribution accounts for the perturbative effect of the
underlying standard model dynamics, in which the amplitude can be calculated
explicitly with the aid of
meson decay constants in the hadron matrix elements.
In principle, we could directly test the standard model parameters
if the short distance effect in the decay is able to be
extracted from the total amplitude, provided the long
distance effect can also be calculated separately. However, the long distance
effects in many $K$
decays dominate the decay amplitudes and thus the short distance contributions
are buried under the overwhelming long distance backgrounds \cite{reviews}.
In such cases, it would be difficult to learn the physics of
the standard model.

Fortunately, there exists a class of $K$ decay modes such as
$K^+\to\pi^+\nu\bar{\nu}$, dominated by the short distance effects
\cite{rein89,hage89,lu94,geng94}. These decays have been playing important
roles for us to understand the structure of weak
interaction with high precision.
In general, they are suppressed by GIM mechanism and the leading
short distance contributions arise from one-loop diagrams,
resulting in that the decay amplitudes involve the CKM matrix
elements and heavy quark masses such as $m_c$ and $m_t$
\cite{bela91,buch94}. Problems concerning the hadron matrix element
persist but can be better managed. The physics that how the quarks
form hadrons, namely the nonperturbative effect, is lumped into
the measured constants $f_\pi$ and $f_K$.
The effect of the short distance contribution can be calculated within
the framework of the standard model and the result is factorized. Since
the decay amplitudes depend explicitly on various weak interaction
parameters such as $V_{td}$
or $m_t$, they can be used to extract these parameters from experimental
data. A common practice, for example, is to plot the decay rate as a
function of $m_t$ to give us some insights on the top quark mass,
which, in turn, could eliminate the uncertainty in the determination
of the CKM parameters.
It is clear that the result relies crucially upon the domination of
the short distance contribution to the decay rate.
In principle, the long distance effect may also contribute to the decay
amplitude at the same order of magnitude.
There is no apparent reason why the effect should be suppressed.
In this paper, however, we will give an explicit calculation in chiral
perturbation theory to illustrate that such effect in the
class of $K\to n\pi\nu\bar{\nu}$ decays is negligible compared with the
short distance one, where $n$ stands for the number of $\pi$ mesons.
Especially we will show that the long distance parts of the decay rates for
the processes $K\to\pi\nu\bar{\nu}$ and $K\to \pi\pi\nu\bar{\nu}$
are at least seven orders of
magnitude smaller than those from the short distance contributions
in the chiral logarithmic approximation.
The origin for the suppression is twofold. The long distance contribution,
which is believed to arise mainly from $u$ quark loop in the underlying
theory, is much smaller compared with the short distance contribution which
contains heavy quark loop contribution\cite{IL}. On the other 
hand, all the long distance contributions in $K\to n\pi\nu\bar{\nu}$ 
modes, calculated within the framework of chiral
perturbation theory, start to receive contributions at $O(P^4)$ as a result
of incompatibility of Lorentz invariance and gauge invariance. The 
dominance of short distance contribution ensures the validity of relating
the decay rate of this sort of processes to the weak interaction parameters.

The article is organized as follows. In Sec. II, we
construct the Lagrangian which
is relevant to the decays of interest. In Sec. III, we calculate the decay
amplitudes of $K\rightarrow n\pi\nu\bar{\nu}$ and give numerical evaluations.
The results of the short and long distance contributions to the decay
rates are compared. We  conclude in Sec. IV.

\section{Chiral Lagrangian Involving $Z^0$}
The processes $K\rightarrow n\pi\nu\bar{\nu}$ are mainly mediated by
$K\rightarrow n\pi Z^0$ and followed by $Z^0\rightarrow\nu\bar{\nu}$. 
We restrict our
discussion on the $Z^0$ mediated diagrams only.
We will not consider the box diagram with two external $W$ bosons as 
intermediate states.
The couplings of $K n\pi Z^0$ are the major concern of this section.
The effect of $Z^0$ can be incorporated into the chiral Lagrangian of mesons
by treating it
as an external gauge field. There are four pieces in the Lagrangian,
which are relevant to the analysis of $K n\pi Z^0$ couplings,
corresponding to the strong interaction, ${\cal L}_2$,
the weak interaction, ${\cal L}^{\Delta S=1}_{2}$,
the Wess-Zumino-Witten (W.Z.W.) anomaly\cite{wess71,witt83}, ${\cal
L}_{WZW}$, and the weak anomaly\cite{bijn92}, ${\cal L}^{\Delta
S=1}_{A}$, respectively. The first two are of
$O(P^2)$ while the last two of $O(P^4)$ in the chiral power counting.
In order to consider
$K\rightarrow n\pi\nu\bar{\nu}$ to $O(P^4)$ consistently, the $O(P^4)$ counter
terms of strong and weak interactions, as well as the loop corrections,
should be all included in the
analysis. However, as is well known, the counter terms contain numerous
unknown
coefficients\cite{kamb90} and it makes the calculation impractical. The
strategy adopted in
the present work is first to decide whether the $O(P^2)$ Lagrangian contributes
to the processes or not. If it does, the amplitude can be
calculated explicitly without the uncertainty arising from the unknown
coefficients and the analysis will
be terminated there since the $O(P^2)$ Lagrangian yields the dominant
contribution. If $O(P^2)$
terms in the Lagrangian do not contribute, then the long distance
contribution is further suppressed. In principle, three kinds
of contributions may enter the
$O(P^4)$ analysis, which are loop corrections, counter terms and
anomalies, respectively. But, as will
be shown in the numerical analysis, the long distance effect in the
decay branching ratio is at least relatively $10^{-7}$ smaller than the
short distance effect\cite{geng94,geng95}, so that exact numerical
evaluations are not necessary. It is sufficient to use the anomalies or the
loop corrections to estimate the orders of magnitude of the long distance
effect.

In the $O(P^2)$ Lagrangian, $Z^0$ is incorporated into the Lagrangian by
gauging derivatives. The $Z^0$ coupling, with the mixing of the hypercharge,
contains both left and right handed currents. The right handed current
has only octet coupling while the left one gets both octet
and singlet ones \cite{lu94,geng94,geng95}. The $U(1)$ symmetry,
resulting in the singlet coupling,
is not the part of the chiral symmetry. We include it by first assuming nonet
symmetry and then use a parameter $\xi$ to indicate the degree of the nonet
symmetry breaking with $\xi=1$ in the nonet symmetry limit. The covariant
derivative reads as
\begin{eqnarray}
D_\mu U & = & \partial_\mu U-i r_\mu U+i U l_\mu
\nonumber\\
& \equiv & \partial_\mu U+{i g \over cos \theta_W} (U Q-{\xi\over 6} U I
-sin^2\theta_W \left[U,Q\right])Z_\mu^0,
\end{eqnarray}
where $Q$ is the quark charge matrix, $Q= diag(2/3,-1/3,-1/3)$, and $U$ is the
nonlinear realization of meson octet
\begin{eqnarray}
U = exp (i \Phi / f_\pi),
\end{eqnarray}
with
\begin{eqnarray}
\Phi = \phi^a \lambda^a = \sqrt{2}  \left(
\begin{array}{ccc}
\pi^0/\sqrt{2}+\eta/\sqrt{6}& \pi^+& K^+ \\
\pi^-& -\pi^0/\sqrt{2}+\eta/\sqrt{6}& K^0 \\
K^-& \bar{K}^0 & -2\eta/\sqrt{6}
\end{array}
\right).
\end{eqnarray}
and $f_{\pi}= 93 MeV$ being the pion decay constant.
By this identification of the covariant derivative, the $O(P^2)$
strong interaction Lagrangian is given by
\begin{eqnarray}
{\cal L}_2
& = & {f_{\pi}^2\over 4} Tr \left[ D_{\mu}U^\dagger
D^{\mu}U+2 B_0 M (U+U^\dagger)\right],
\end{eqnarray}
where $M=diag(m_u, m_d, m_s)$ is the quark mass matrix.
Note that the covariant derivative defined here is different from that in
Ref. [4]. In fact, the left and right handed currents were
switched in Ref. [4].
It is not consistent with the required chiral transformation property
\cite{geng95}. The coefficient
of the mass term is determined by the ratios of meson and quark masses
\begin{eqnarray}
B_0 = {m_K^2\over m_u+m_s} = {m_\pi^2\over m_u+m_d} =
{3 m_\eta^2\over m_u+m_d+4 m_s}.
\end{eqnarray}
The $O(P^2)$ weak interaction Lagrangian is given by
\begin{eqnarray}
{\cal L}^{\Delta S=1}_2
& = & G_8 f_{\pi}^4 Tr \lambda_6 D_{\mu}U^\dagger
D^{\mu}U,
\end{eqnarray}
with $G_8 \approx 9.1\times 10^{-6} GeV^{-2}$ determined from
$K\rightarrow\pi\pi$.
The relevant part of W.Z.W. anomaly has only one gauge boson
coupling \begin{eqnarray}
{\cal L}_{WZW}& = &
- {1\over 16 \pi^2}
\varepsilon^{\mu\nu\alpha\beta}Tr
(\Sigma^L_{\mu}\Sigma^L_{\nu}\Sigma^L_{\alpha}l_{\beta}
- \Sigma^R_{\mu}\Sigma^R_{\nu}\Sigma^R_{\alpha}r_{\beta}),
\end{eqnarray}
where
$\Sigma^L_\mu = U^\dagger \partial_\mu U$ and
$\Sigma^R_\mu = U \partial_\mu U^\dagger$.
The direct weak anomaly\cite{bijn92} contains four unknown parameters $a_i$
$(i=1\cdots 4)$ and, explicitly, one has,
\begin{eqnarray}
{\cal L}^{\Delta S=1}_{A} & = & {G_8 f_{\pi}^2 \over 16 \pi^2}\left\{ 2a_1 i
\varepsilon^{\mu\nu\alpha\beta}Tr\lambda_6 L_{\mu}
Tr L_{\nu}L_{\alpha}L_{\beta}\right. \nonumber\\
& & + a_2 Tr \lambda_6 [U^{\dagger}\tilde{F}^{\mu\nu}_R U,
L_{\mu}L_{\nu}] + 3a_3Tr\lambda_6 L_{\mu} Tr (\tilde{F}^{\mu\nu}_L +
U^{\dagger}\tilde{F}^{\mu\nu}_R U)L_{\nu} \nonumber\\
& & \left. + a_4 Tr\lambda_6 L_{\mu} Tr (\tilde{F}^{\mu\nu}_L -
U^{\dagger}\tilde{F}^{\mu\nu}_R U)L_{\nu}\right\}.
\end{eqnarray}
These unknown parameters are believed to be of order one. The
fields in Eq. (8) are defined as
\begin{eqnarray}
& & L_\mu  =  i U^\dagger D_\mu U\; , R_\mu  =  i U D_\mu
U^\dagger\nonumber\\
& & F^L_{\mu\nu}  =  \partial_\mu l_\nu - \partial_\nu l_\mu\; ,
F^R_{\mu\nu}  =  \partial_\mu r_\nu - \partial_\nu r_\mu\nonumber\\
& & \tilde{F}_{L,R}^{\mu\nu}  =
\varepsilon^{\mu\nu\rho\sigma}F^{L,R}_{\rho\sigma}.
\end{eqnarray}
Since $K$ could mix with $\pi$ through a weak transition, it is cumbersome
to use the octet fields defined in Eq. (3) to calculate the amplitudes.
The new basis which simultaneously
diagonalizes ${\cal L}_2$ and ${\cal L}^{\Delta S=1}_{2}$\cite{ecke88}
is given by
\begin{eqnarray}
\pi^+ & \rightarrow &\pi^+ - {2 m_K^2 f_\pi^2 G_8 \over m_K^2 - m_\pi^2}
K^+\nonumber\\
K^+ & \rightarrow & K^+ + {2 m_\pi^2 f_\pi^2 G_8^* \over m_K^2 - m_\pi^2}
\pi^+\nonumber\\
\pi^0 & \rightarrow & \pi^0 + {\sqrt{2} m_K^2 f_\pi^2 \over m_K^2
- m_\pi^2} ( G_8 K^0 + G_8^* \bar{K}^0)\nonumber\\
K^0 &\rightarrow & K^0 - {\sqrt{2} m_\pi^2 f_\pi^2 G_8^* \over m_K^2 - m_\pi^2}
\pi^0 + \sqrt{{2 \over 3}}{m_\eta^2 f_\pi^2 G_8^* \over m_\eta^2 - m_K^2}
\eta\nonumber\\
\eta & \rightarrow & \eta - \sqrt{{2 \over 3}} {m_K^2 f_\pi^2 \over m_\eta^2 -
m_K^2} (G_8 K^0 + G_8^* \bar{K}^0)\,.
\end{eqnarray}
In this transformed basis,
the vertices $K^+\pi^+$, $K_L\pi^0$, $K_L\pi^0 Z^0$ and $K^+\pi^+ Z^0$
are eliminated and the numbers of Feynman diagrams for the
processes of interest are reduced substantially.

\section{Amplitudes and Branching Ratios}
The phase space allowed modes for $K$ decaying to
$n\pi\nu\bar{\nu}$ are $\pi\nu\bar{\nu}$, $\pi\pi\nu\bar{\nu}$
and $\pi\pi\pi\nu\bar{\nu}$, i.e.,  $n\leq 3$.
It is easily seen that processes involving only
neutral particles receive no contribution from the Lagrangian.
The long distance contributions in
$K_L\to\pi^0\nu\bar{\nu}\,,\ \pi^0\pi^0\nu\bar{\nu}$ and
$\pi^0\pi^0\pi^0\nu\bar{\nu}$ are therefore trivial and we shall not
include these modes in the remaining discussions of the paper.
The amplitudes and branching ratios of the rest decays of
$K\rightarrow n\pi\nu\bar{\nu}$ are analyzed
with the aforementioned strategy as follows.

\subsection*{$K^+\rightarrow\pi^+\nu\bar{\nu}$}
The long distance contribution of this mode has recently been
studied in various approaches \cite{rein89,lu94,geng95}.
It is found that neither the $O(P^2)$ Lagrangian nor the anomaly Lagrangian
contributes to this process\cite{geng95}. So the amplitude is at most of
$O(P^4)$. The loop contribution\cite{geng95} is given by
\begin{eqnarray}
A(K^+\rightarrow\pi^+\nu\bar{\nu})  & = &
-{i\alpha G_8 (1-2\sin^2 \theta_W)\over 64\pi M_Z^2\sin^2 \theta_W\cos^2
\theta_W} J(m_K^2) (P_K + P_\pi)^{\mu}\bar{\nu}\gamma_{\mu}(1-\gamma_5)\nu ,
\end{eqnarray}
where the loop function $J(m^2)$ is defined as
\begin{eqnarray}
J(m^2) & = & {1 \over i\pi^2} \int d^nq {1\over q^2 - m^2}\nonumber\\
     & = & m^2(\Delta - \ln { m^2\over 4\pi^2 f_{\pi}^2}).
\end{eqnarray}
The divergent part in Eq. (12) is given by
\begin{eqnarray}
\Delta & = & {2\over \epsilon}- \gamma -\ln \pi + 1,
\end{eqnarray}
where $\gamma$ is the Euler number and $\epsilon=4-n$.
The decay rate can be evaluated analytically and it is found to be
\begin{eqnarray}
\Gamma(K^+\rightarrow\pi^+\nu\bar{\nu}) & = &
{\alpha^2 G_8^2 m_K^5 (1-2\sin^2 \theta_W)^2
\over 2^{19}\pi^5 M_Z^4 \sin^4\theta_W \cos^4 \theta_W}\nonumber\\
& & \cdot (1-8r_\pi+8r_\pi^3-r_\pi^4 -12r^2_\pi \ln r_\pi)|J(m_K^2)|^2,
\end{eqnarray}
with $r_\pi=m_\pi^2/m_K^2$.
The long distance contribution gives rise to the branching ratio
\begin{eqnarray}
Br(K^+\rightarrow\pi^+\nu\bar{\nu}) & \sim & 7.7\cdot 10^{-18}\,,
\end{eqnarray}
which is roughly $10^{-7}$ smaller than that of the short distance
contribution\cite{hage89,bela91}. 
We note that our result in Eq. (15)
is different from that of Refs. \cite{rein89} and \cite{lu94}.
Although our work is analyzed within the same framework, namely chiral 
perturbation theory as in \cite{lu94}, 
we find that the tree level amplitude of $K^+\pi^+Z^0$ vanishes 
identically as shown in Sec. II, which is only true in the limit of
the large $N_c$ in \cite{lu94}.
This difference may arise from the different identification of 
left-handed and right-handed currents \cite{geng95}.
However, we could not be able to find the reason of the discrepancy with 
Ref. \cite{rein89}. Further study on this issue is needed.

\subsection*{$K_L\rightarrow\pi^+\pi^-\nu\bar{\nu}$}
This mode receives no contribution from the $O(P^2)$ Lagrangian. The
amplitudes arising from the anomaly Lagrangian can be written as
\begin{eqnarray}
A(K_L\rightarrow\pi^+\pi^-\nu\bar{\nu}) & = &-{i\alpha G_8\over 4\pi f_\pi
M_Z^2\sin^2\theta_W\cos^2\theta_W}
\bar{\nu}\gamma_\mu(1-\gamma_5)\nu\nonumber\\
&& \cdot\left[6a_1-6a_3-2a_4+2\sin^2\theta_W(a_2+2a_4)
-\xi\right]
\varepsilon^{\mu\nu\alpha\beta}
P_{K\nu}P_{\pi^+\alpha}P_{\pi^-\beta}.
\end{eqnarray}
The corresponding differential decay rate can be evaluated analytically
\begin{eqnarray}
{d^3\Gamma\over ds_\pi ds_\nu d\cos\theta_\pi} & = &
{\alpha^2 G_8^2 \sigma_\pi^3X^3\sin^2\theta_\pi s_\pi s_\nu\over
2^{15}\pi^7f_\pi^2 M_Z^4 m_K^3 \sin^4\theta_W\cos^4 \theta_W}
\nonumber\\
&& \cdot\left[6a_1-6a_3-2a_4
+2\sin^2\theta_W(a_2+2a_4) - \xi\right]^2,
\end{eqnarray}
where
\begin{eqnarray}
s_{\pi}&=&(P_{\pi^+}+P_{\pi^-})^2\;,\
s_{\nu}\:=\:(P_{\nu}+P_{\bar{\nu}})^2\;,\
\sigma_{\pi}\:=\:(1-4m_{\pi}^2/s_{\pi})^{1/2} ,\nonumber\\
X&=&\left\{\left[{1\over 2}(m_K^2-s_{\pi}-s_{\nu})\right]^2
-s_{\pi}s_{\nu}\right\}^{1/2}.
\end{eqnarray}
It leads to a branching ratio
\begin{eqnarray}
Br(K_L\rightarrow\pi^+\pi^-\nu\bar{\nu}) & = & 4.81\cdot10^{-20}[6a_1 - 6a_3
-2a_4 + 2\sin^2\theta_W(a_2+2a_4)-\xi]^2\,,
\end{eqnarray}
which is roughly $10^{-7}$ smaller compared with the short distance
contribution\cite{geng94}.

\subsection*{$K^+\rightarrow\pi^+\pi^0\nu\bar{\nu}$}
Like the previous decay mode,
$K^+\rightarrow\pi^+\pi^0\nu\bar{\nu}$ receives only
contribution from the anomaly Lagrangian. The amplitude is given by
\begin{eqnarray}
A(K^+\rightarrow\pi^+\pi^0\nu\bar{\nu}) & = & -{i\alpha G_8\over 4\pi
f_{\pi}M_Z^2\sin^2\theta_W\cos^2\theta_W} \bar{\nu}\gamma_{\mu}(1-\gamma_5)\nu
\nonumber\\
& & \cdot\left[1+3a_3+a_4-\sin^2\theta_W(2-3a_2+6a_3)
-\xi\right]
\varepsilon^{\mu\nu\alpha\beta}
P_{K\nu}P_{\pi^+\alpha}P_{\pi^0\beta}.
\end{eqnarray}
The differential decay rate\cite{bijn92a,bijn94} is found to be
\begin{eqnarray}
{d^3\Gamma\over ds_\pi ds_\nu d\cos\theta_\pi} & = &
{\alpha^2 G_8^2 \sigma_\pi^3 X^3\sin^2\theta_\pi s_\pi s_\nu\over
2^{15}\pi^7f_\pi^2 M_Z^4m_K^3\sin^4\theta_W\cos^4\theta_W}
\nonumber\\
&&
\cdot\left[1+3a_3+a_4-\sin^2\theta_W(2-3a_2+6a_3)-\xi\right]^2,
\end{eqnarray}
where
\begin{eqnarray}
s_{\pi}\:=\:(P_{\pi^+}+P_{\pi^0})^2\;,\
s_{\nu}\:=\:(P_{\nu}+P_{\bar{\nu}})^2\;,\
\sigma_{\pi}\:=\:(1-(m_{\pi^+}+m_{\pi^0})^2/s_{\pi})^{1/2} ,\nonumber\\
X\:=\:\left\{\left[{1\over 2}(m_K^2-s_{\pi}-s_{\nu})\right]^2
-s_{\pi}s_{\nu}\right\}^{1/2}
\end{eqnarray}
The branching ratio is obtained as
\begin{eqnarray}
Br(K^+\rightarrow\pi^+\pi^0\nu\bar{\nu}) & = & 3.44\cdot10^{-19}[1 +
3a_3 +a_4 - \sin^2\theta_W(2-3a_2+6a_3)-\xi]^2
\end{eqnarray}
which is about one order larger than that of
$K_L\rightarrow\pi^+\pi^-\nu\bar{\nu}$.
Since $K^+\rightarrow\pi^+\pi^0\nu\bar{\nu}$
is related to $K_L\rightarrow\pi^+\pi^-\nu\bar{\nu}$ by isospin symmetry for
both the long and short distance contributions, the relative
suppression between the long and short distance effects
should be roughly the same.

\subsection*{$K\rightarrow 3\pi\nu\bar{\nu}$}
There are three modes $K_L\rightarrow\pi^+\pi^-\pi^0\nu\bar{\nu}$,
$K^+\rightarrow\pi^+\pi^0\pi^0\nu\bar{\nu}$
and $K^+\rightarrow\pi^+\pi^+\pi^-\nu\bar{\nu}$ receive contributions from
the $O(P^2)$ Lagrangian. The processes go through $K\rightarrow 3\pi$ and then
emit $Z^0$ from one of the charged meson involved. They are basically
internal bremsstrahlung type of processes. Since the leading
contribution is of $O(P^2)$,
the suppression is not very strong. Unfortunately, there are no short distance
calculations available at the moment because of the smallness of the
rate due to the phase space suppression and we only list the decay
amplitudes by the long distance contributions for the sake of
completeness. They are obtained as follows
\begin{eqnarray}
A(K_L\rightarrow \pi^+\pi^-\pi^0\nu\bar{\nu}) & = &
-{i\pi\alpha (1-2\sin^2\theta_W)G_8\over
M_Z^2\sin^2\theta_W\cos^2\theta_W} (m_K^2-2P_KP_{\pi^0}) \nonumber\\
&& \cdot\left[{P_{\pi^+}^\mu\over P_Z(P_Z+2P_{\pi^+})}-
{P_{\pi^-}^\mu\over P_Z(P_Z+2 P_{\pi^-})}\right]
\bar{\nu}\gamma_{\mu}(1-\gamma_5)\nu
\end{eqnarray}

\begin{eqnarray}
A(K^+\rightarrow \pi^+\pi^0\pi^0\nu\bar{\nu}) & = &
{i\pi\alpha (1-2\sin^2\theta_W) G_8\over M_Z^2\sin^2\theta_W\cos^2\theta_W}
{m_\pi^2+2P_{\pi^0}P'_{\pi^0}\over P_Z(P_Z+2P_{\pi^+})}P_{\pi^+}^\mu
\bar{\nu}\gamma_{\mu}(1-\gamma_5)\nu
\end{eqnarray}

\begin{eqnarray}
A(K^+\rightarrow \pi^+\pi^+\pi^-\nu\bar{\nu}) & = &
{i\pi\alpha(1-2\sin^2\theta_W) G_8\over
M_Z^2\sin^2\theta_W\cos^2\theta_W}\nonumber\\
&& \cdot\left[{m_K^2+m_\pi^2-2P_KP'_{\pi^+}+2P'_{\pi^+}P_{\pi^-}\over
P_Z(P_Z+2P_{\pi^+})} P_{\pi^+}^\mu\right.\nonumber\\
&& \left.
+{m_K^2+m_\pi^2-2P_KP_{\pi^+}+2P_{\pi^+}P_{\pi^-}\over
P_Z(P_Z+2P'_{\pi^+})}{P'}_{\pi^+}^{\mu} \right.\nonumber\\
&& \left.
-{2m_K^2-P_KP_{\pi^+}-P_KP'_{\pi^+}\over P_Z(P_Z+2P_{\pi^-})}P_{\pi^-}^\mu
\right]\bar{\nu}\gamma_{\mu}(1-\gamma_5)\nu\,,
\end{eqnarray}
where $P'_{\pi^{0(+)}}$ represents the momentum of the second
$\pi^0(\pi^+)$ in the relevant mode and $P_Z$ is equal to
the difference between the momenta of the kaon and three pions or
the sum of the two neutrino momenta.

\section{Conclusion}
We have calculated the long distance contributions to the
decays of $K\rightarrow n\pi\nu\bar{\nu}$
within the framework of chiral perturbation theory. For the processes
with one or two pions in the
final states, the long distance effect is highly suppressed relative to the
short distance by a factor of $10^{-7}$. 
We remark that the estimates of the decay rates have been done by 
the replacement of the divergent loop function of Eq. (12) by its finite 
part, the so-called chiral logarithmic piece, which is a rather crude 
approximation. 
We have also neglected all $O(P^4)$ counter terms 
in the chiral Lagrangian.
Moreover, 
we have ignored the contribution from the box diagram with two W-boson 
intermediate states, which might be large as shown in Ref. [2].
Therefore, our results are subject some uncertainties. 
However, these uncertainties should not change the conclusion that the 
long-distance contributions to $K\to n\pi\nu\bar{\nu}$ are negligible 
compared to that from the short-distance parts.
Extraction of standard model parameters from these modes suffers no 
uncertainty from the contamination of the long distance effect.

\acknowledgments
This work is supported by the National Science Council of the
ROC under contract numbers NSC84-2122-M-007-013 and NSC84-2112-M-007-041.

\end{document}